\documentclass[aps,prc,twocolumn,showpacs,amsmath,amssymb,nofootinbib]{revtex4-2}
\usepackage{graphicx}
\usepackage{color}
\usepackage{times}
\usepackage{inputenc}
\usepackage{bm}
\usepackage{ulem}
\usepackage{multirow}
\usepackage{float}
\usepackage{url}
\usepackage{natbib}
\usepackage{mathrsfs}
\usepackage{physics}
\usepackage[table,xcdraw]{xcolor}
\usepackage{booktabs}
\usepackage{comment}
\usepackage[colorlinks=true,citecolor=blue,urlcolor=blue,linkcolor=blue]{hyperref}
\begin{document}
\title{Constraining the Surface Curvature of an Anisotropic Neutron Star}
\author{H. C. Das$^{1,2}$}
\email{harish.d@iopb.res.in}
\author{Jeet Amrit Pattnaik$^{3}$}
\author{S. K. Patra$^{1,2}$}
\affiliation{\it $^{1}$Institute of Physics, Sachivalaya Marg, Bhubaneswar 751005, India}
\affiliation{\it $^{2}$Homi Bhabha National Institute, Training School Complex, Anushakti Nagar, Mumbai 400094, India}
\affiliation{\it $^{3}$Department of Physics, Siksha $'O'$ Anusandhan, Deemed to be University, Bhubaneswar -751030, India}
\date{\today}
\begin{abstract}
The anisotropy of pressure arises due to the various complex phenomena that happen inside the neutron star (NS). In this study, we calculate the degree of anisotropy inside the NS using the scalar pressure anisotropy model. Macroscopic properties such as mass, radius, compactness, redshift, tidal deformability, the moment of inertia, and surface curvature (SC) are computed for the anisotropic NS with the equation of states spanning from relativistic to nonrelativistic cases. The variation of SC as the functions of the above-mentioned quantities are computed by changing the degree of anisotropy. Pressure anisotropy has significant effects on the magnitude of SC. The relations between the canonical SC$-\Lambda$ and SC$-\bar{I}$ are studied. From the GW170817 tidal deformability data, we constraints the magnitude of SC are found to be SC$_{1.4}(10^{14}) = 3.44_{-1.0}^{+0.4}, 2.85_{-1.20}^{+0.62}, 2.52_{-1.02}^{+0.61}$ for $\lambda_{\rm BL}  = 0.0, 1.0$, and $2.0$ respectively.
\end{abstract}
\maketitle
\section{Introduction} 
One of the most difficult tasks is to examine the internal structure of neutron stars (NS), which requires knowledge from many different branches of physics. Due to its intricate internal structure and powerful gravitational pull, we haven't yet developed a comprehensive theoretical understanding of this object. Several exotic processes are happening inside the NS, which makes the system more complex. In addition, we consider the pressure anisotropy inside the star, which is a realistic phenomenon. There are different sources of pressure anisotropy inside the NS, such as a strong magnetic field \cite{Yazadjiev_2012, Cardall_2001, Ioka_2004, Ciolfi_2010, Ciolfi_2013, Frieben_2012, Pili_2014, Bucciantini_2015}, pion condensation \cite{Carter_1998}, phase transitions \cite{Carter_1998}, core crystallization \cite{Nelmes_2012}, and a superfluid core \cite{Kippenhahn_1990, NKGb_1997, Heselberg_2000}. For details regarding the sources of anisotropy and its significance, see Ref. \cite{Herrera_1997}.

In literature, few anisotropic models have been developed, such as Bowers-Liang (BL) \cite{Bowers_1974}, Horvat {\it et al.} \cite{Horvat_2010}, and Cosenza {\it et al.} \cite{Cosenza_1981} models. The BL model is based on the assumption that (i) the anisotropy quadratically disappears at the origin, (ii)  it varies non-linearly on radial pressure, and (iii) the anisotropy is gravitationally driven. According to the hypothesis by Horvat {\it et al.} \cite{Horvat_2010}, the anisotropy results from the quasi-local equation as described in Ref. \cite{Horvat_2010}. According to several research, the limits of the anisotropic parameter are $-2\leq \lambda_{\rm BL}\leq +2$ for BL model \cite{Silva_2015}, and $-2\leq \lambda_{\rm H}\leq +2$ in the Horvat model \cite{Doneva_2012}. In this scenario, the BL model is taken into account to calculate the various NS properties. 

Numerous research clarified the impacts of anisotropic pressure on the macroscopic characteristics of compact objects, such as their mass, radius, the moment of inertia, tidal deformability, and non-radial oscillation \cite{Bowers_1974, Horvat_2010, Cosenza_1981, Biswas_2019, Sulaksono_2015, Sulaksono_2020, Rahmansyah_2020, Rahmansyah_2021, Silva_2015, Deb_2021}. The magnitudes of macroscopic characteristics often rise when the magnitude of the anisotropy parameter increases, and vice versa. In this study, we intend to determine the surface curvature (SC) as a function of different NS observables for an anisotropic NS. It is an important quantity to measure the curvature at the surface of the star. Also, the SC of the NS is approximately $10^{14}$ orders of magnitude more than our solar counterpart \cite{Kazim_2014}. In the near future, we may be able to measure such curvature using the Shapiro delay measurement technique \cite{Rosi_2015}.   

Here, we explore the relations between SC-tidal deformability and SC-moment of inertia. Universal/approximate relations play a great role to determine the quantity that is not accessible to detect if there exists a relation between them. For example, the moment of inertia-tidal Love number and quadrupole ($I-$Love$-Q$) relation \cite{Yagi_2015}. If we detect the Love number, we can calculate the other two using that relation. Different types of relations exist in the literature \cite{Yagi_2015, Haskell_2014, Das_ILC_2022, Landry_2018, Kumar_2019, Bikram_2023}. In our previous calculations, we calculated the $I-$Love$-C$ (where $C$ is the compactness of a star) for different degrees of anisotropy \cite{Das_ILC_2022}. We have put a limit on the moment of inertia of the anisotropic NS using various observational data. In this work, we want to find the relations between SC$-\Lambda$ and SC$-\bar{I}$ of the anisotropic NS. However, the approximate relations are different than the universal relations. Sometimes approximate relations are helpful in determining the behavior of the observables. An approximate relation is used when it is not feasible or practical to obtain the complete and accurate nature of the data. For example, in data mining, it may be acceptable to have a relation with missing values or errors if the goal is to identify patterns or trends in the data. In the present scenario, the relations between SC, tidal deformability, and the moment of inertia become weaker in case of the maximum-mass star. However, the relations are strong between them in the case of canonical stars, which will be very helpful in future aspects.

Hence, to explore those properties of the NS, one needs EOSs which can be taken either from the relativistic mean-field model (RMF) or the Skyrme-Hartree-Fock (SHF). In this calculation, we have taken RMF, SHF, and density-dependent (DD-RMF) EOSs as described in our previous study \cite{Das_ILC_2022}. Here, we choose the unified EOSs to explore the anisotropic NS properties. The EOSs which we have taken in this calculation produce the mass of the NS $\sim 2 \ M_\odot$. The manuscript is categorized as follows. After a brief introduction, we provide the formalism to calculate the pressure anisotropy inside the star. The expressions and results for SC, the moment of inertia, and tidal deformability are given in Sec-\ref{sec:form}. The numerical results are given in Sec-\ref{sec:RD}. Finally, we enumerate the summary and conclusion in Sec-\ref{sec:sc}. 
\section{Anisotropic configurations}
\label{sec:form}
For an isotropic star with the perfect fluid condition, the Maxwell stress-energy tensor is defined as \cite{Walecka_74}
\begin{eqnarray}
    T_{\mu\nu} = ({\cal{E}}+P)u_\mu u_\nu + P g_{\mu\nu},
    \label{eq:tmunu}
\end{eqnarray}
where ${\cal{E}}$, $P$, and $u_\mu$ are the energy density, pressure, and 4-velocity of the perfect fluid, respectively. However, various phenomena inside the star make the pressure anisotropic in the tangential direction. That means the magnitudes of radial pressure ($P_r$) differ from the tangential pressure ($P_t$). For that case, the stress-energy tensor is defined as ~\cite{Doneva_2012, Silva_2015, Estevez_2018}
\begin{eqnarray}
    T_{\mu\nu} = ({\cal{E}}+P_t)u_\mu u_\nu + (P_r-P_t) k_\mu k_\nu + P_t g_{\mu\nu},
    \label{eq:tmunu_aniso}
\end{eqnarray}
where $k_\mu$ is the unit radial vector ($k^\mu k_\mu = 1$) with $u^\mu k_\mu = 0$. The Tolman-Oppenheimer-Volkoff (TOV) equations for an anisotropic star is defined as  ~\cite{Doneva_2012}
\begin{eqnarray}
    \frac{dP_r}{dr}=-\frac{\left( {\cal E} + P_r \right)\left(m + 4\pi r^3 P_r \right)}{r\left(r -2m\right)} +\frac{2}{r} (P_t - P_r)\,,
    \label{tov1:eps}
\end{eqnarray}
\begin{eqnarray}
    \frac{dm}{dr}=4\pi r^{2}{{\cal E}}\,,
    \label{tov2:eps}
\end{eqnarray}
where $\sigma=P_t-P_r$ is the anisotropy parameter. The `$m$' is the enclosed mass correspond to radius $r$. To solve these TOV equations, we need two separate EOSs for radial and tangential pressure. We use the various RMF, DD-RMF, and SHF equation of states for radial pressure $P_r ({\cal E})$, and the BL model for transverse pressure $P_t$ \cite{Bowers_1974}. The TOV equations can be solved using the boundary conditions $r=0, m=0, P_r=P_c$, and $r=R, m = M, \ {\rm and} \ P_r=0$ for a particular choice of anisotropy. 
\subsection{BL Model}
According to the BL model, the local anisotropy inside the star is based on two main assumptions (i) The anisotropy quadratically disappears at the origin, and (ii) At least a part of the anisotropy is gravitationally driven, and it is nonlinear with the pressure. The anisotropy pressure along the tangential direction $P_t$ is defined as \cite{Bowers_1974}
\begin{eqnarray}
    \label{Anisotropy_eos}
    P_t = P_r + \frac{\lambda_{\rm BL}}{3} \frac{({\cal E}+3P_r)({\cal E} + P_r)r^2}{1-2m/r} \,,
\end{eqnarray}
where the factor $\lambda_{\rm BL}$ measures the degree of anisotropy in the fluid. We assume that the value of $\lambda_{\rm BL}$ is in the range -2 to +2 following the Silva {\it et al.} \cite{Silva_2015}. This range is purely model dependent. One can take different ranges and put constraints using different observational data. Like, Silva {\it et al.} \cite{Silva_2015} has taken $-2<\lambda_{\rm BL}<+2$. Biswas {\it et al.} \cite{Biswas_2019} has constrained the $\lambda_{\rm BL}$ limit using GW170817 tidal deformability data. In our previous study \cite{Das_ILC_2022}, we found that $\lambda_{\rm BL}=+2$ satisfied the secondary mass limit of the GW190814 event. Also, the canonical radius for an anisotropic star satisfied the NICER constraints for $-2<\lambda_{\rm BL}<+2$. Therefore, in this study, we vary the degree of anisotropicity in the range of $-2$ to $+2$.
\section{Surface Curvature}
\label{fcurvature}
In this Sub-Sec., we adopt the curvature quantity from the Refs. \cite{Kazim_2014, Das_2021}. The Kretschmann scalar (full contraction of the Riemann tensor) is defined as 
\begin{eqnarray}
{\cal{K}}(r)&\equiv&\sqrt{{\cal{R}}^{\mu\nu\rho\sigma}{\cal{R}}_{\mu\nu\rho\sigma}}
\nonumber\\
&=& \bigg[(8\pi)^2\left \{3{\cal{E}}^2(r)+3P^2(r)+2P(r){\cal{E}}(r)\right\}
\nonumber\\
&&
-\frac{128{\cal{E}}(r)m(r)}{r^3} +\frac{48m^2(r)}{r^6}\bigg]^{1/2},
\label{KS}
\end{eqnarray}
At the surface, $m\rightarrow M$ as $r \rightarrow R$. Except for the last term, all other terms vanish outside the star because they depend on ${\cal{E}}_{\rm NS}(r)$, and  $P_{\rm NS}(r)$, which are zero outside the star. But, there is a non-vanishing component of the Riemann tensor that does not vanish; $\tensor{{\cal R}}{^1_{010}}=-\frac{2M}{R^3}=- \xi$, even in the outside of the star \cite{Kazim_2014, Xiao_2015}. Therefore, the Riemann tensor is a more relevant quantity to measure the curvature of the stars. Kretschmann scalar is the square root of the full contraction of the Riemann tensor. The vacuum value for both $\cal{K}$ is $\frac{4\sqrt{3}M}{R^3}$, one can easily  see from Eq. (\ref{KS}). Therefore, one can take $\cal K$ reasonable measures for the curvature within the star. The SC is defined as the ratio of curvature at the surface of the NS ${\cal{K}}(R)$ to the curvature of the Sun ${\cal{K}}_\odot$, SC $={\cal{K}}(R)/{\cal{K}}_\odot$. This ratio ${\cal{K}}(R)/{\cal{K}}_\odot\approx10^{14}$ i.e, the NS curvature is $10^{14}$ times more than the Sun.
\section{Moment of Inertia}
\label{moi}
For a slowly rotating NS, the system's equilibrium position can be obtained by solving Einstein's equation in the Hartle-Throne metric as ~\cite{Hartle_1967, Hartle_1968, Hartle_1973} 
\begin{align}
    ds^2  = & -e^{2\nu} \ dt^2 + e^{2\lambda} \ dr + r^2 \ (d\theta^2 +\sin^2\theta d\phi^2)
    \\ \nonumber
    & - 2\omega(r)r^2\sin^2\theta \ dt \ d\phi.
\end{align}
The MI of the slowly rotating anisotropic NS is calculated in Ref. ~\cite{Sulaksono_2020} 
\begin{align}
    I &= \frac{8\pi}{3}\int_0^R \frac{r^5J\Tilde{\omega}}{r- 2M}({\cal{E}}+P)\left[1+\frac{\sigma}{{\cal{E}}+P}\right] \, dr,
    \label{eq:MI}
\end{align}
where $\Tilde{\omega}=\Bar{\omega}/\Omega$, where $\Bar{\omega}$ is the frame dragging angular frequency, $\Bar{\omega} = \Omega-\omega(r)$. $J$ is defined as $e^{-\nu}(1-2m/r)^{1/2}$. 
\begin{figure}
    \centering
    \includegraphics[width=0.5\textwidth]{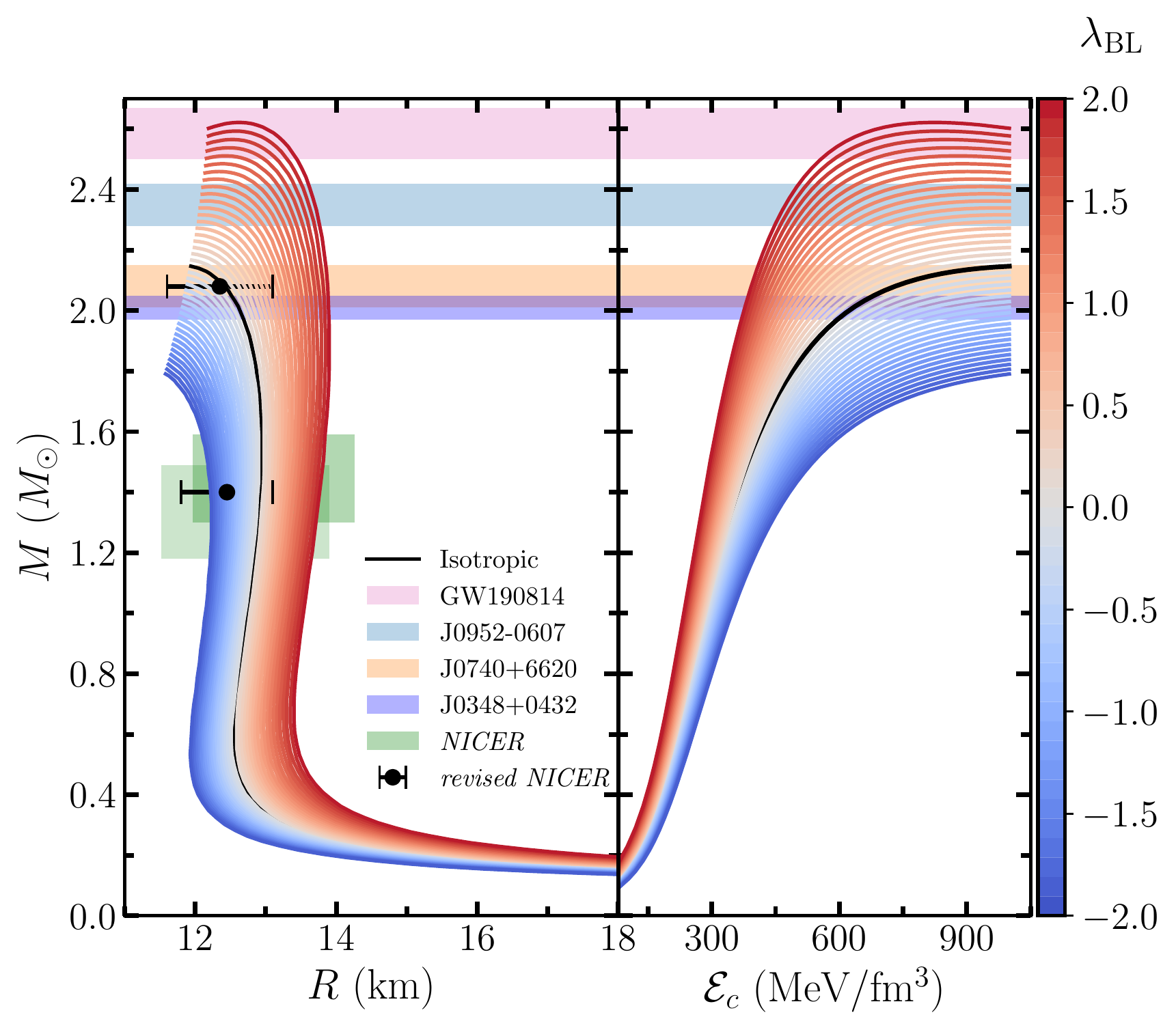}
    \caption{{\it Left:} Mass-radius profiles for anisotropic NS with $-2.0<\lambda_{\rm BL}<+2.0$ for IOPB-I EOS. Different color bands signify the masses of the NS observed from the various pulsars, such as PSR J0348+0432 \cite{Antoniadis_2013}, PSR J0740+6620 \cite{Fonseca_2021}, heaviest pulsars J0952-0607 \cite{Romani_2022}, and GW190814 \cite{RAbbott_2020}. The NICER results are shown with two green boxes from two different analyses ~\cite{Miller_2019, Riley_2019}. The revised NICER results are also shown for the canonical star and $2.08 \ M_\odot$ (red horizontal error bars) given by  Miller {\it et al.} ~\cite{Miller_2021}. {\it Right:} Variation of mass as a function of central density. The black line represents the isotropic star.}
    \label{fig:mr}
\end{figure}
\section{Tidal Deformability}
\label{sec3}
The shape of the NS is deformed when it is present in the external field ($\epsilon_{ij}$) of its companion. Hence the stars develop the quadrupole moment ($Q_{ij}$), which is linearly dependent on the tidal field and is defined as ~\cite{Hinderer_2008, Hinderer_2009}
\begin{eqnarray}
    Q_{ij}=-\lambda \epsilon_{ij}\,,
\end{eqnarray}
where $\lambda$ is defined as the tidal deformability of a star. It has relation to the dimensionless tidal Love number $k_2$ as $\lambda = \frac{2}{3} k_2 R^5$, where $R$ is the radius of the star. The dimensionless tidal deformability is defined as $\Lambda=\lambda/M^5=2 k_2/3 C^5$, where $C$ is the compactness of the star defined as $M/R$.

To determine $k_2$, we use the linear perturbation in the Throne and Campolattaro metric \cite{Throne_1967}. We have solved the Einstein equation and obtained the following second-order differential equation for the anisotropic star \cite{Biswas_2019}
\begin{align}
H^{''} &+ H^{'} \bigg[\frac{2}{r} + e^{\lambda} \left(\frac{2m(r)}{r^2} + 4 \pi r (P - {\cal E})\right)\bigg] 
\nonumber \\
&
+ H \left[4\pi e^{\lambda} \left(4 {\cal E} + 8P + \frac{{\cal E} + P}{dP_t/d{\cal E}}(1+c_s^2)\right) -\frac{6 e^{\lambda}}{r^2} - {\nu^\prime}^2\right] 
\nonumber \\
&
= 0\,.
\end{align}
The term $dP_t/d{\cal E}$ represents the change of $P_t$ (see Eq. (\ref{Anisotropy_eos}) for the $P_t$) with respect to energy density for a fixed value of $\lambda_{\rm BL}$.

The internal and external solutions to the perturbed variable $H$ at the star's surface can be matched to get the tidal Love number \cite{Damour_2009, Hinderer_2008}. The value of the tidal Love number can be calculated using the $y_2$, and compactness parameter $C$ is defined as ~\cite{Hinderer_2008, Hinderer_2009, DasBig_2021}
\begin{align}
    k_2 &= \, \frac{8}{5} C^5 (1-2C)^2 \big[ 2(y_2-1)C - y_2 + 2 \big]
    \nonumber \\ &
    \times \Big\{ 2C \big[ 4(y_2+1)C^4 + 2(3y_2-2)C^3 - 2(11y_2-13)C^2 
    \nonumber \\ &
    + 3(5y_2-8)C - 3(y_2-2) \big]+ 3(1-2C)^2 
    \nonumber \\ &
    \times \big[ 2(y_2-1)C-y_2+2 \big] \log(1-2C) \Big\}^{-1} \, , 
    \label{eq:k2}
\end{align}
where $y_2$ depends on the surface value of $H$ and its derivative 
\begin{equation}
    y_2 = \frac{rH^{'}}{H}\Big|_R.
\end{equation}
\begin{figure}
    \centering
    \includegraphics[width=0.5\textwidth]{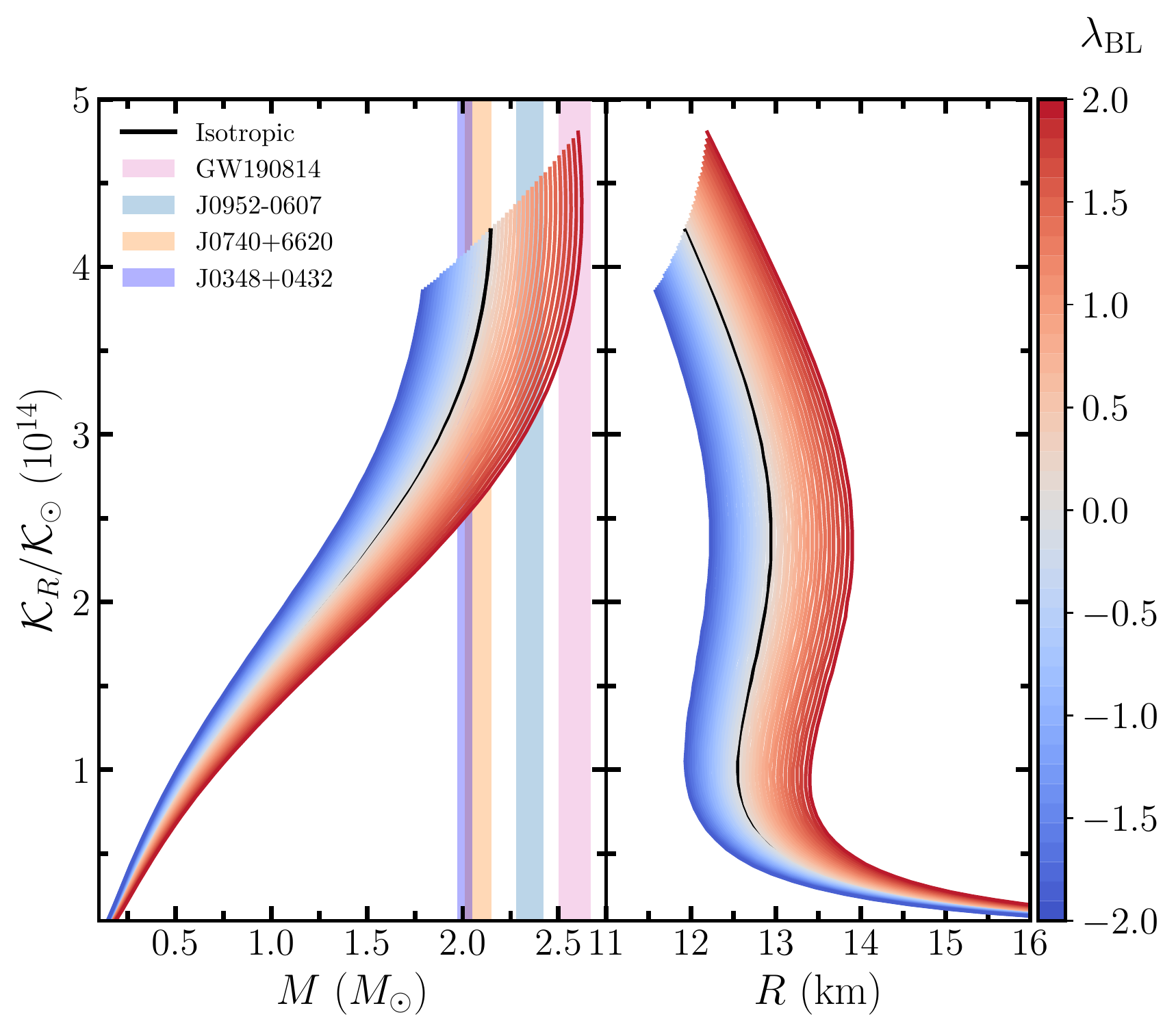}
    \caption{{\it Left:} SC as a function of mass for anisotropic NS with $-2.0<\lambda_{\rm BL}<+2.0$ for IOPB-I equation of state. {\it Right:} SC as a function of radius.}
    \label{fig:kmr}
\end{figure}
\begin{figure}
    \centering
    \includegraphics[width=0.5\textwidth]{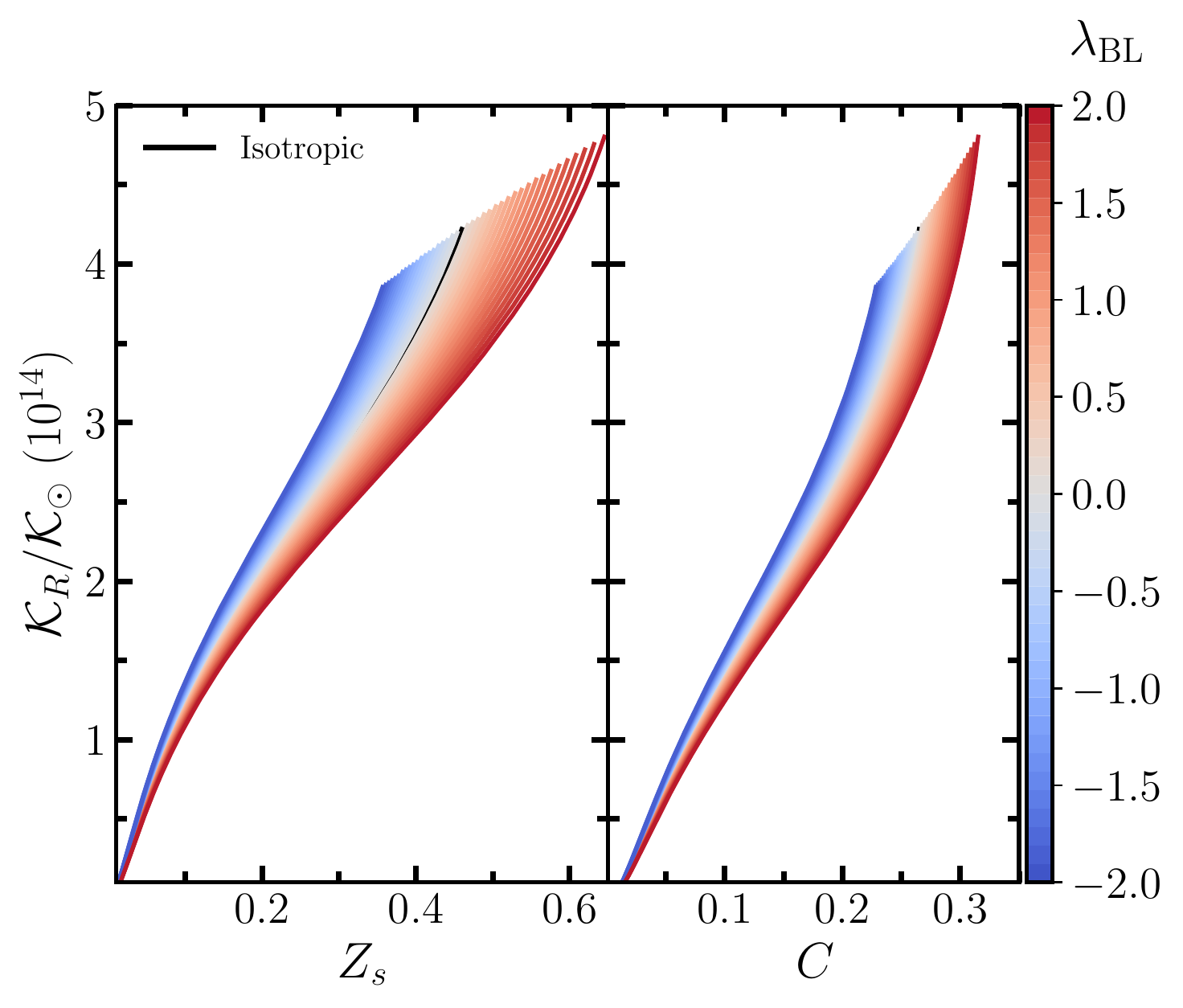}
    \caption{{\it Left:} SC as a function of red-shift for anisotropic NS with $-2.0<\lambda_{\rm BL}<+2.0$ for IOPB-I equation of state. {\it Right:} SC as a function of compactness.}
    \label{fig:kzc}
\end{figure}
\section{Results and Discussions}
\label{sec:RD}
\subsection{Mass-Radius relations}
For IOPB-I equation of state, the mass-radius profiles of the anisotropic NS are solved for various values of BL, as illustrated in Fig. \ref{fig:mr}. The magnitude of the maximum masses and their corresponding radii increases with increasing positive values of $\lambda_{\rm BL}$, and vice versa. Different observational data, including x-ray, NICER, and GW, may constrain the degree of anisotropy within the NS. Recently, it is discovered that PSR J0952-0607, the fastest and heaviest Galactic NS in the Milky Way disc, has a mass $M = 2.35\pm 0.17 \, M_\odot$ \cite{Romani_2022}. Additionally, we can put this restriction to control the degree of anisotropy.

The GW190814 event sparked a debate about whether the secondary component is the lightest black hole or the heaviest neutron star \cite{RAbbott_2020}. To explain this behavior, several models have already been proposed in the literature \cite{Roupas_2021, Fattoyev_2020, Huang_2020, Lim_2021, DasPRD_2021}. However, Roupas {\it et al.} Ref. \cite{Roupas_2021} asserted that the secondary component might be an anisotropic NS. Hence, we investigated the mass-radius diagram with the secondary component mass limit $M = 2.50 -2.67 \ M_\odot$ to see if it replicated the limit for anisotropic stars in the BL model. The mass $2.50–2.67 \ M_\odot$ can be reproduced for $\lambda_{\rm BL} = 1.8–2$, although those values of $\lambda_{\rm BL}$ do not adhere to the revised NICER limits \cite{Miller_2021}.
\begin{figure}
    \centering
\includegraphics[width=0.5\textwidth]{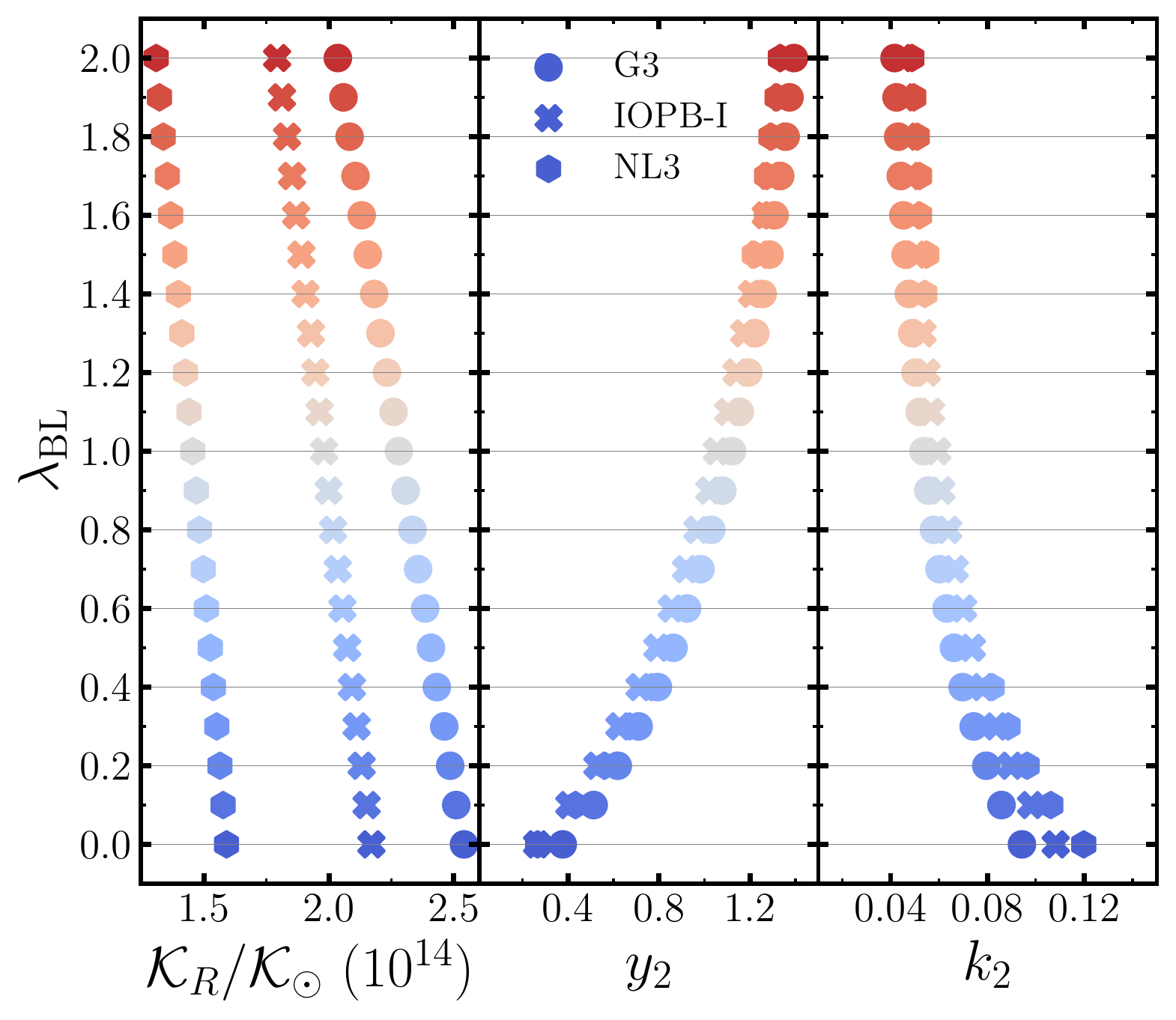}
    \caption{SC, $y_2$, and $k_2$ with the variation of $0.0<\lambda_{\rm BL}<+2.0$ for canonical anisotropic NS for G3 (soft), IOPB-I (moderate stiff), and NL3 (stiff) EOSs.}
    \label{fig:Ky2k2_cano}
\end{figure}
\begin{figure}
    \centering
    \includegraphics[width=0.5\textwidth]{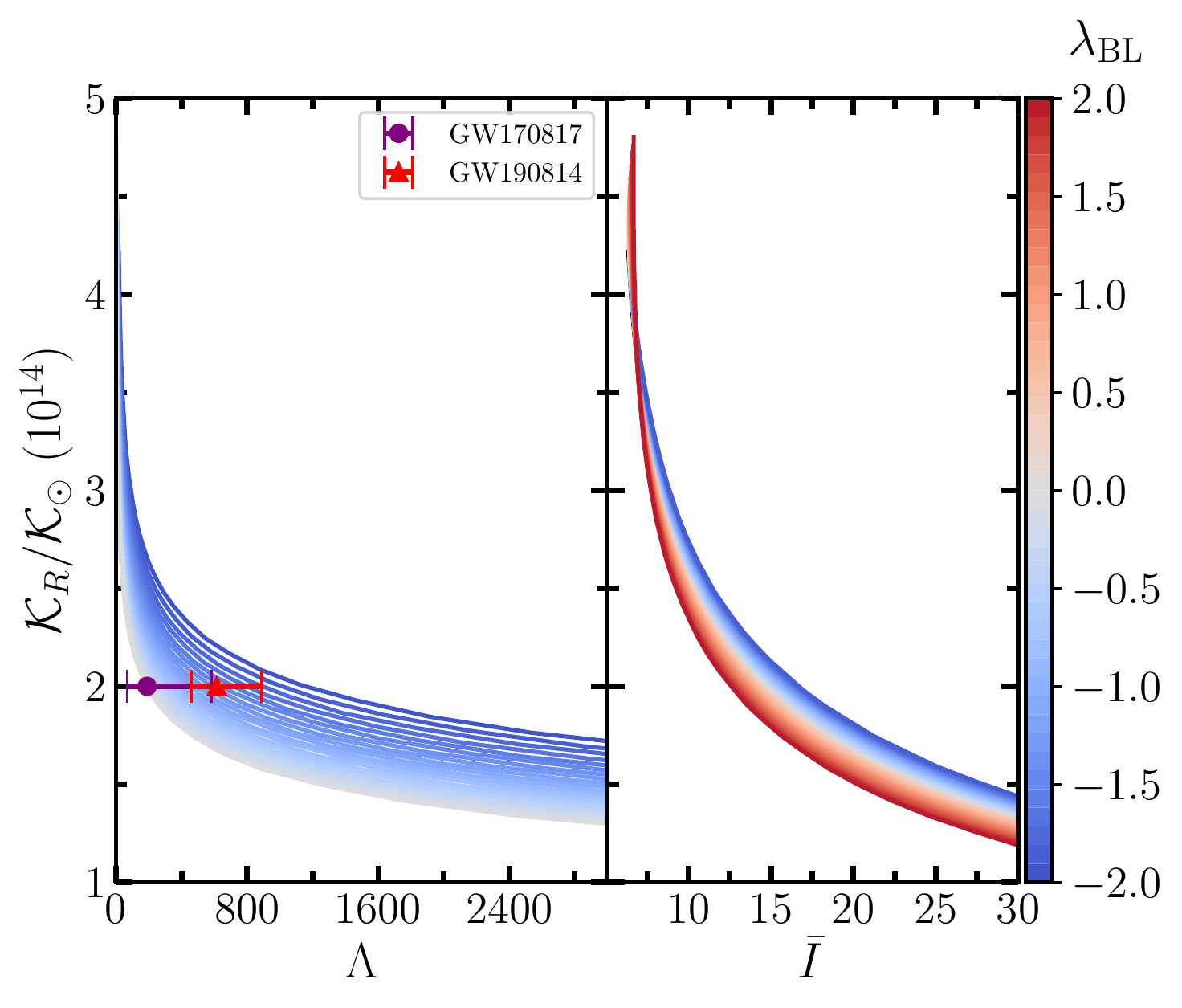}
    \caption{{\it Left:} SC as a function of $\Lambda$ for anisotropic NS with $0.0<\lambda_{\rm BL}<+2.0$ for IOPB-I equation of state. The error bars are the observational constraints given by LIGO/Virgo events GW170817 (NS-NS merger) ~\cite{Abbott_2017} and GW190814 (assuming BH-NS merger) ~\cite{RAbbott_2020}}. {\it Right:} SC as a function of $\bar{I}$ for $-2.0<\lambda_{\rm BL}<+2.0$.
    \label{fig:kli}
\end{figure}
\subsection{Surface Curvatures}
We calculate the SC with the IOPB-I equation of state by changing the value of $\lambda_{\rm BL}$, which is shown in Fig. \ref{fig:kmr}. The magnitude of SC increases both with the mass of the star as well as with the positive value of $\lambda_{\rm BL}$ and vice-versa. This is due to the fact that the variation of positive values $\lambda_{\rm BL}$ enhances the magnitude of the tangential pressure of the star, which supports large gravitational mass. In the previous section, we explained that the SC is directly proportional to the mass of the star. Therefore, with increasing the mass of the star, the SC increases. However, the negative values of $\lambda_{\rm BL}$ follow the opposite trend, which is clearly seen in the figure. Also, the radial variation of SC is depicted in the same figure. The variation in SC mainly follows the  $1/R^3$ trend. The magnitude of the radius increases towards the surface, which decreases SC. However, it enhances its magnitude near the center. Various observational bands are the data from which one can constrain the magnitude of curvature, as well as the amount of anisotropy inside the star \cite{Das_ILC_2022}. Recently, the  gravity-field curvature has been measured using the three conjugated atom interferometers \cite{Rosi_2015}. In the future, we may be able to measure the SC of the NS by either using interferometers or Shapiro delay measurements. 

The parametric dependence of the variation of SC has been discussed in our previous work \cite{Das_2021}. The softer EOS, such as G3, provides more SC in comparison to IOPB-I (see Fig. 5 of Ref. \cite{Das_2021}). This is due to the fact that the magnitude of both $M$ and $R$ decreases for softer EOS in comparison to stiffer ones. Hence, the quantity $M/R^3$ for soft EOS is more. Also, it increases more for dark matter (DM) admixed star due to the fact that with the addition of DM, the EOS becomes softer \cite{Das_2020, Das_2021, DasMNRAS_2021, Dasfmode_2021, DasPRD_2021}.  

In a similar way, we show the variations of SC with red-shift $\left(Z_s = \frac{1}{\sqrt{1-2C}}-1\right)$ and compactness in Fig. \ref{fig:kzc} for IOPB-I equation of state as a representative case. The magnitude of SC increases with increasing both for $Z_s$ and $C$. Since the redshift is a function of compactness only, therefore, the trend of changing SC is the same as compactness. However, the magnitude of both $Z_s$ and $C$ increases with increasing the value of $\lambda_{\rm BL}$. From different observational data, we can put constrain on the degree of anisotropy inside the star. Until now, just one value of $Z_s=0.35$ from the analysis of stacked bursts in Ref. \cite{Cottam_2002} has been reported in the low-mass x-ray binary EXO 0748-676, which is also discarded by the subsequent observation \cite{Cottam_2008}.

The variation of SC, $y_2$, and $k_2$ by varying only positive values of $\lambda_{\rm BL}$ are shown in Fig. \ref{fig:Ky2k2_cano} for the canonical star. In this case, we take three different types of EOSs, such as G3 (soft) \cite{Parmar_2022}, IOPB-I (moderately stiff) \cite{Parmar_2022}, and NL3 (stiff) \cite{Lalazissis_1997}, for representative cases. The magnitude of SC is larger for G3 in comparison to IOPB-I and NL3. However, the magnitude of $y_2$ and $k_2$ are almost the same for the three EOSs with the variation of $\lambda_{\rm BL}$. This simply indicates the model-independent nature of $y_2$ and $k_2$ for different values of $\lambda_{\rm BL}$ for the canonical anisotropic NS.

The variation of SC both for dimensionless tidal deformability and the moment of inertia for the IOPB-I equation of state with different values of $\lambda_{\rm BL}$ is shown in Fig. \ref{fig:kli}. We observe that the magnitude of SC decreases with increasing either $\Lambda$ or $\bar{I}$. Also, the relations seem to be correlated for a fixed $\lambda_{\rm BL}$ as well as the mass. The relation becomes tighter in the case of SC$-\bar{I}$ as compared to SC$-\Lambda$. One can put a direct constraint on the SC from the various observational data if any relation exists between them. Therefore, in the next sub-section, we try to get the functional relations between them by changing the degree of anisotropy.
\begin{figure*}
    \centering
    \includegraphics[width=0.5\textwidth]{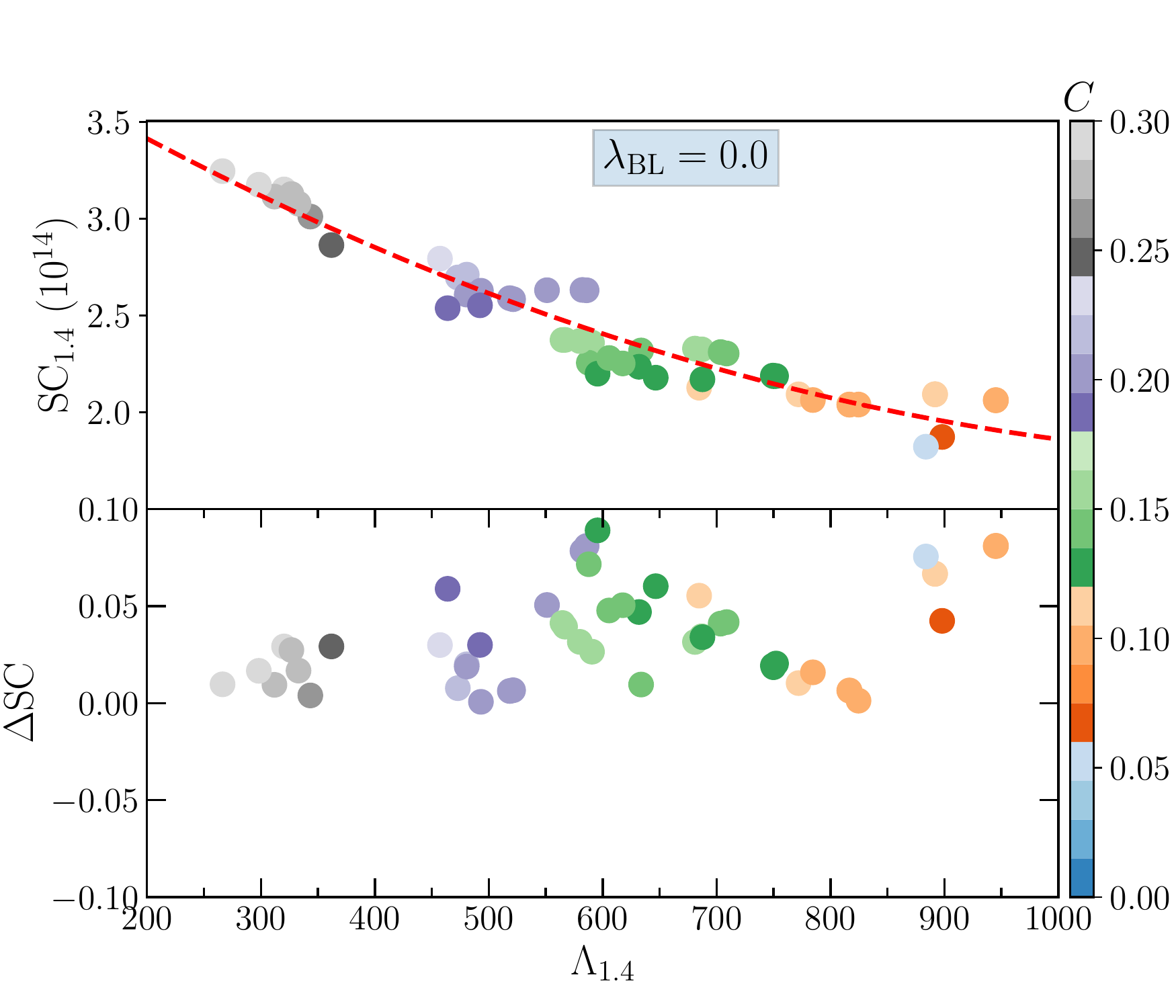}%
    \includegraphics[width=0.5\textwidth]{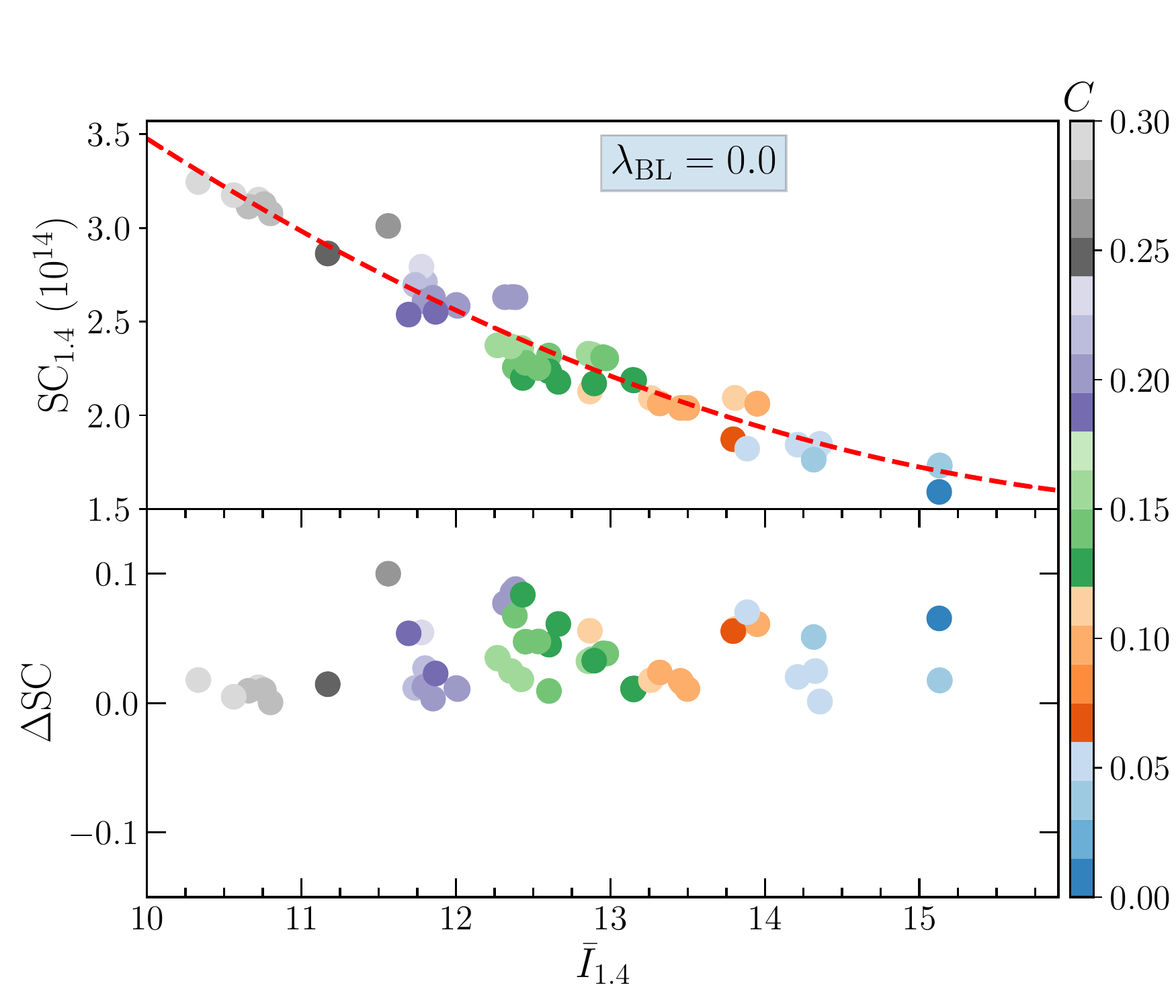}
    \includegraphics[width=0.5\textwidth]{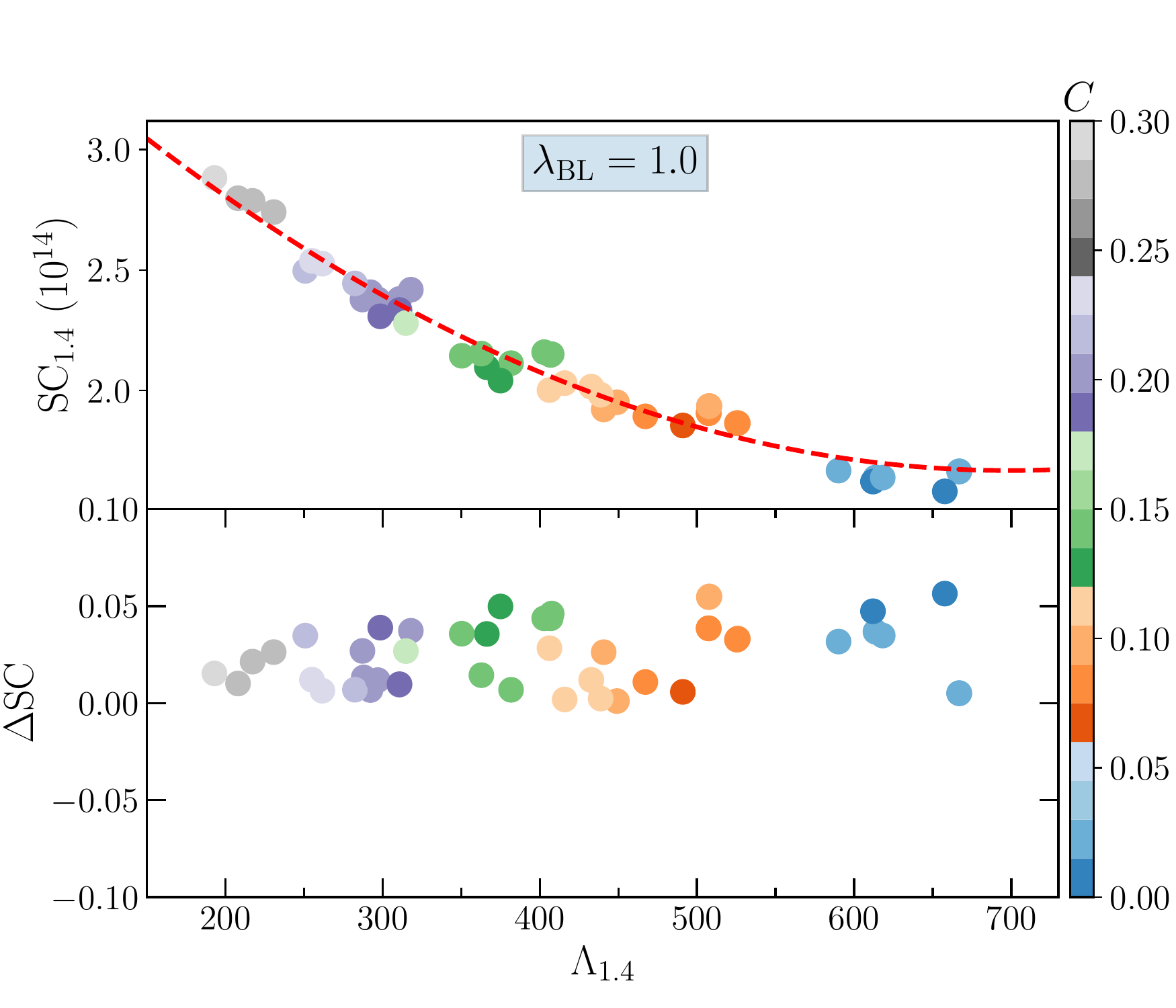}%
    \includegraphics[width=0.5\textwidth]{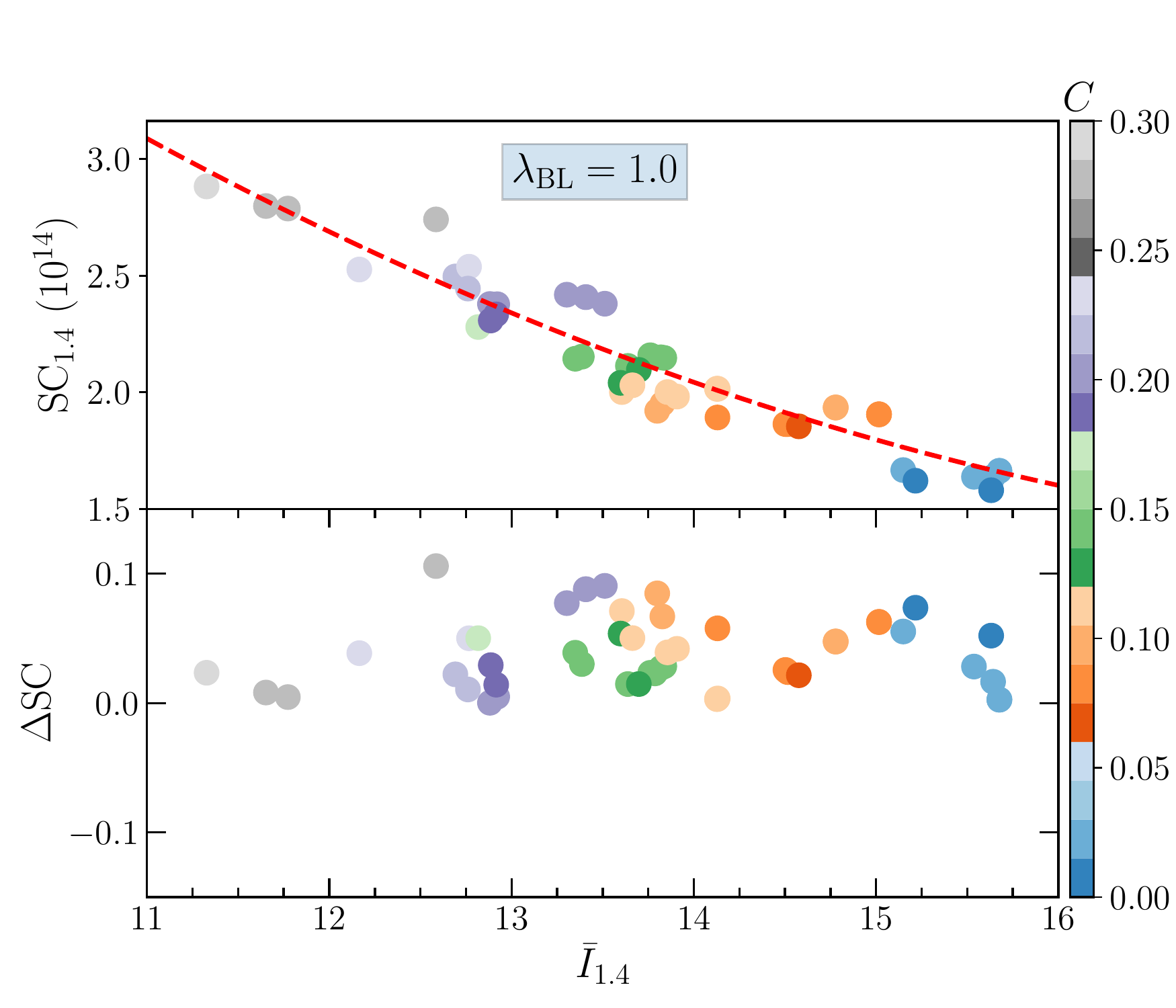}
    \includegraphics[width=0.5\textwidth]{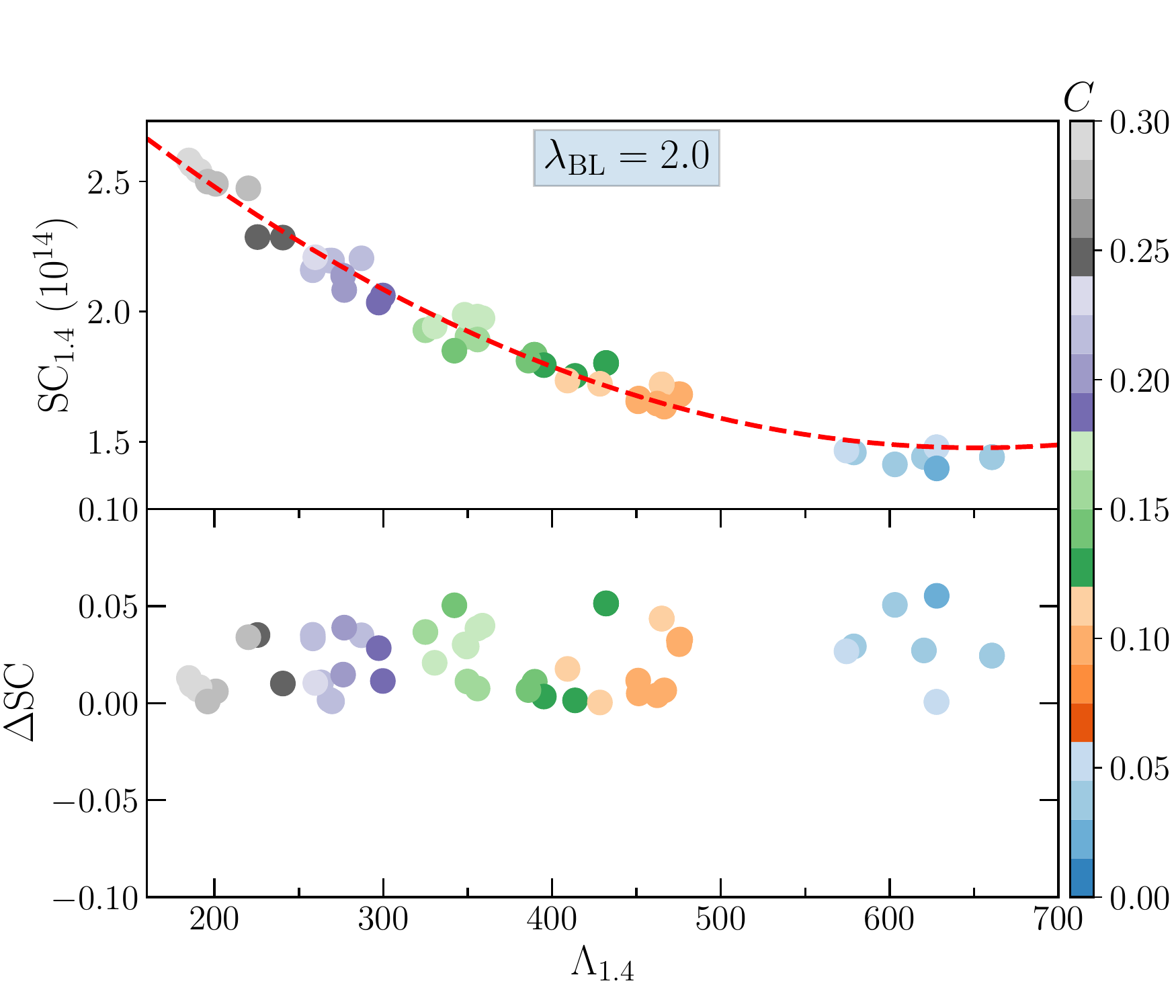}%
    \includegraphics[width=0.5\textwidth]{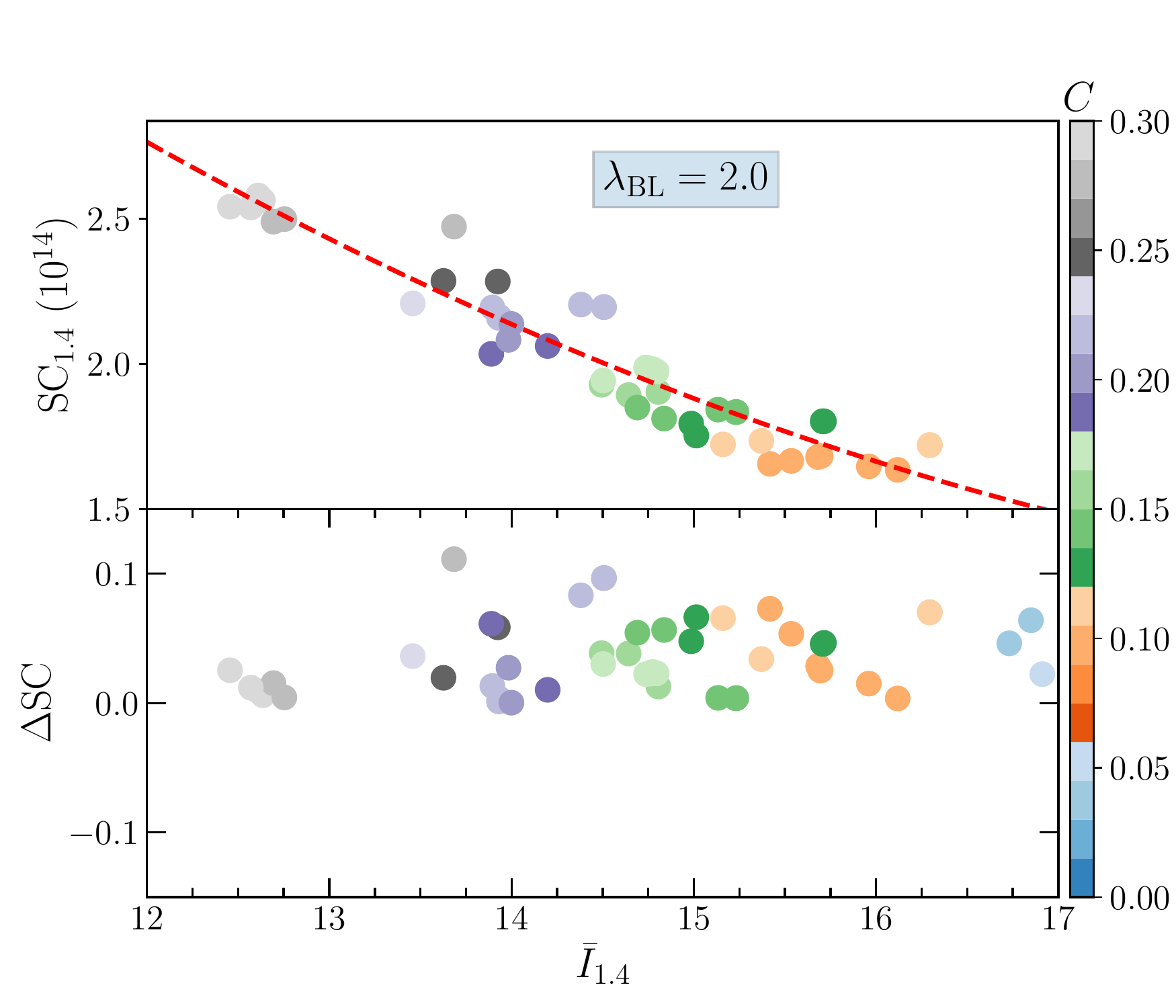}
    \caption{Relations between SC$_{1.4}-\Lambda_{1.4}$ (left), and SC$_{1.4}-\bar{I}_{1.4}$ (right) by varying anisotropy parameter $\lambda_{\rm BL}$. The lower panels in each figure are for the residual of the fitting using Eq. (\ref{eq:residual_fit}).}
    \label{fig:sc_l_i}
\end{figure*}
\begin{table}
    \centering
    \caption{The fitting coefficients are listed for SC$_{1.4}-\Lambda_{1.4}$, and SC$_{1.4}-\bar{I}_{1.4}$ relations with $\lambda_{\rm BL} = 0.0, 1.0, $ and $2.0$ respectively.}
    \label{tab:fit_coefficients}
    \renewcommand{\arraystretch}{1.5}
    \scalebox{0.81}{
        \begin{tabular}{llllllll}
            \hline \hline
            \multicolumn{4}{l}{\hspace{2.5cm}SC$_{1.4}-\Lambda_{1.4}$}
            &\multicolumn{4}{l}{\hspace{2.5cm}SC$_{1.4}-\bar{I}_{1.4}$} \\ \hline
            \multirow{2}{*}{$\lambda_{\rm BL} =$} &
            \multirow{2}{*}{0.0} &
            \multirow{2}{*}{1.0} &
            \multirow{2}{*}{2.0} &
            \multirow{2}{*}{$\lambda_{\rm BL} =$} &
            \multirow{2}{*}{0.0} &
            \multirow{2}{*}{1.0} &
            \multirow{2}{*}{2.0} \\ 
            &  &  &  &                      &  &  &     \\ \hline
            $a_0   =$  & $4.0908$ & $3.9029$ & $3.5629$  & $b_0 =$ &$12.3498$  &$10.8343$  &$9.7856$ \\ 
            $a_1 (10^{-3}) =$  & $-3.6762$ & $-6.3974$ & $-6.3987$  & $b_1 =$ &$-1.2453$  &$-0.9840$ & $-0.8173$   \\ 
            $a_2 (10^{-6}) =$  & $ 1.4463$& $4.57661$ & $4.9067$  & $b_2 =$ &$0.0358$  &$0.0254$  & $0.0194$ \\ \hline \hline
    \end{tabular}}
\end{table}
\subsection{Relations between SC, $\Lambda$, and $\bar{I}$ for a canonical star}
Various universal relations have already been proposed, such as $I-$Love$-Q$, $I-$Love-$C$, etc., both for static as well as rotating NS. In this case, we want to calculate the approximate/universal functional relations between SC$-\Lambda$ and SC$-I$ for an anisotropic star.

The canonical value for SC$-\Lambda$ with spans of almost 60 EOSs are calculated and shown in left panels of Fig. \ref{fig:sc_l_i} for the anisotropic star cases for $\lambda_{\rm BL} = 0.0, 1.0$, and $2.0$. In this calculation, we take only positive values of $\lambda_{\rm BL}$. This is because the higher negative values of $\lambda_{\rm BL}$ give unphysical solutions of $\Lambda$, and lower negative values don't change the magnitude of $\Lambda$ significantly, as discussed in Refs. \cite{Biswas_2019, Das_ILC_2022}. The calculated data are fitted with the approximate formula using the least-square method \cite{Das_ILC_2022}
\begin{equation}
    {\rm SC}_{1.4} = \sum_{n=0}^2 a_n (\Lambda_{1.4})^n,
    \label{eq:fit_equation}
\end{equation}
where $a_n$ is the fitting coefficient given in Table \ref{tab:fit_coefficients}. The lower panel of the figure represents the residuals for the fit calculated using the formula \cite{Das_ILC_2022}
\begin{align}
    \Delta{\rm SC} = \frac{|{\rm SC}-{\rm SC}_{\rm fit}|}{{\rm SC}_{\rm fit}}.
    \label{eq:residual_fit}
\end{align}

Another relation between SC$_{1.4}-\bar{I}_{1.4}$ is also shown in the right panels of Fig. \ref{fig:sc_l_i} for $\lambda_{\rm BL} = 0.0, 1.0$, and $2.0$. We use the same fitting formula as given in Eq. (\ref{eq:fit_equation}), but the coefficients are different as given in Table \ref{tab:fit_coefficients} for $b_n$. We also checked for this calculation for the maximum mass star; however, the relationships were found to be weak in comparison to the canonical mass. Hence, in this study, we mainly focus on the relations between SC$-\Lambda$ and SC$-\bar{I}$ for the canonical star.

From these approximate relations, one can put constrain on the SC of the NS with the help of observational data of $\Lambda$ and $\bar{I}$. The tidal deformability limit given by GW170817 from the merger of two binary NS is $\Lambda_{1.4} = 190_{-120}^{+390}$. We obtain the value of SC for the canonical star SC$_{1.4} (10^{14}) = 3.44_{-1.0}^{+0.4}, 2.85_{-1.2}^{+0.62}, 2.52_{-1.02}^{+0.61}$ for $\lambda_{\rm BL}  = 0.0, 1.0$, and $2.0$ respectively. To our knowledge, till now, we don't have any observational data for the moment of inertia of the NS. However, various theoretical studies have been dedicated to constraining the value of the moment of inertia using various observational data. From these limits, one can also constrain the magnitude of surface curvature. 
\section{Summary and Conclusions}
\label{sec:sc}
In this study, we have explored various properties of anisotropic NS within the scalar anisotropic BL model. The magnitude of various macroscopic properties changes with the degree of anisotropy. This is due to the fact that the pressure difference between the radial and transverse components has an additional contribution to balance the hydrostatic equilibrium, and it is purely model-dependent. Any anisotropic models should satisfy the different perfect fluid conditions as mentioned in Refs. \cite{Das_ILC_2022}. 

We have calculated various macroscopic properties such as mass, radius, tidal deformability, the moment of inertia, and surface curvature for a variety of EOSs spanning from relativistic to non-relativistic cases. We observed that the magnitude of maximum mass and its corresponding radius increases (decreases) with the positive (negative) values of $\lambda_{\rm BL}$. However, the magnitude of canonical tidal deformability value decreases with positive values of $\lambda_{\rm BL}$ and vice-versa. Again, the effect of $\lambda_{\rm BL}$ on $y_2$ and $k_2$ are almost similar for all those considered sets, showing the model-independent nature for the anisotropic NS. The surface curvature of the NS is a crucial quantity which is $10^{14}$ orders of magnitude more than the sun. Here, we mainly focus on the variation of SC with some NS properties by varying degrees of anisotropy inside the NS. The SC plays a great role to describe the curvature made by the star in the space-time metric. Using different observational data, we can put stringent constrain on its magnitude if we can find any relationship between known observables such as mass, radius, and tidal deformability. In another way, we can put constraints on the direct observational data such as Shapiro delay measurement. Therefore, in this study, we have explored the relationships between SC and other quantities for the anisotropic NSs. 

The magnitude of SC as a function of mass increases (decreases) with the positive (negative) value of $\lambda_{\rm BL}$. This is because the SC directly depends on mass. However, the radial variation depends on ($1/R^3$), which means, at the core, the SC is more in comparison to the surface of the star. Since almost all mass of the NS is mainly concentrated in the core. In a similar fashion, the SC follows the same trends for the variation of redshift and compactness with the degree of anisotropy. The SC as a function of either tidal deformability or the moment of inertia decreases by varying anisotropicity. From these relations, we observed that there might be some relationship between the SC$-\Lambda$ and SC$-\bar{I}$. 

We have found the approximate relations between SC$-\Lambda$ and SC$-\bar{I}$ for the canonical star. However, the universality becomes weaker for the maximum mass of the star. From the relation between SC$_{1.4}-\Lambda_{1.4}$, we put a constraint on the canonical SC using the GW170817 tidal deformability data. The obtained values are SC$_{1.4} (10^{14}) = 3.44_{-1.0}^{+0.4}, 2.85_{-1.2}^{+0.62}, \ {\rm and} \ 2.52_{-1.02}^{+0.61}$ for $\lambda_{\rm BL}  = 0.0, 1.0$, and $2.0$ respectively. However, till now, we don't have any observational data on the dimensionless moment of inertia. Maybe in the future, by the observation of double pulsars, we can find its value and put stringent constraints on the surface curvature as well as the degree of anisotropy.
\section*{Acknowledgement} 
JAP acknowledges to Institute Of Physics (IOP), Bhubaneswar, for providing the computer facilities during the work. SERB (Project Nos. CRG/2019/002691) partly reinforces this work.
\bibliography{curvature}
\bibliographystyle{apsrev4-2}
\end{document}